\documentclass[twoside]{article}%
\usepackage{amssymb}
\usepackage{amsfonts}
\usepackage{amsmath}
\usepackage{graphicx}%
\providecommand{\U}[1]{\protect\rule{.1in}{.1in}}
\newtheorem{theorem}{Theorem}

\newtheorem{corollary}[theorem]{Corollary}

\newtheorem{lemma}[theorem]{Lemma}

\newenvironment{proof}[1][Proof]{\noindent\textbf{#1.} }{\ \rule{0.5em}{0.5em}}
\begin{document}

\title{\textbf{Perturbational Blowup Solutions to the 1-dimensional Compressible
Euler Equations}}
\author{M\textsc{anwai Yuen\thanks{E-mail address: nevetsyuen@hotmail.com }}\\\textit{Department of Applied Mathematics, The Hong Kong Polytechnic
University,}\\\textit{Hung Hom, Kowloon, Hong Kong}}
\date{Revised 09-Dec-2010}
\maketitle

\begin{abstract}
We study the construction of analytical non-radially solutions for the
1-dimensional compressible adiabatic Euler equations in this article. We could
design the perturbational method to construct a new class of analytical
solutions. In details, we perturb the linear velocity:%
\begin{equation}
u=c(t)x+b(t)
\end{equation}
and substitute it into the compressible Euler equations. By comparing the
coefficients of the polynomial, we could deduce the corresponding functional
differential system of $(c(t),b(t),\rho^{\gamma-1}(0,t)).$ Then by skillfully
applying the Hubble's transformation:
\begin{equation}
c(t)=\frac{\dot{a}(t)}{a(t)},
\end{equation}
the functional differential equations can be simplified to be the system of
$(a(t),b(t),\rho^{\gamma-1}(0,t))$. After proving the existence of the
corresponding ordinary differential equations, a new class of blowup or global
solutions can be shown. Here, our results fully cover the previous known ones
by choosing $b(t)=0$.



\end{abstract}

\section{Introduction}

The isentropic compressible Euler equations can be written in the following
form:
\begin{equation}
\left\{
\begin{array}
[c]{rl}%
{\normalsize \rho}_{t}{\normalsize +\nabla\cdot(\rho u)} & {\normalsize =}%
{\normalsize 0}\\
{\normalsize (\rho u)}_{t}{\normalsize +\nabla\cdot(\rho u\otimes u)+\nabla P}
& {\normalsize =0.}%
\end{array}
\right.  \label{Euler}%
\end{equation}
As usual, $\rho=\rho(x,t)$ and $u=u(x,t)\in\mathbf{R}^{N}$ are the density and
the velocity respectively with $x=(x_{1},$ $x_{2},$ $...,$ $x_{N})\in R^{N}$.
For some fixed $K>0$, we have a $\gamma$-law on the pressure $P=P(\rho)$, i.e.%
\begin{equation}
{\normalsize P}\left(  \rho\right)  {\normalsize =K\rho}^{\gamma}
\label{gamma}%
\end{equation}
which is a common hypothesis. The constant $\gamma=c_{P}/c_{v}>1$, where
$c_{P}$, $c_{v}$\ are the specific heats per unit mass under constant pressure
and constant volume respectively, is the ratio of the specific heats, that is,
the adiabatic exponent. For the solutions in radially symmetry:%
\begin{equation}
\rho(x,t)=\rho(r,t)\text{ and }u(x,t)=\frac{x}{r}V(r,t)=:\frac{x}{r}V
\end{equation}
where the radial $r=\sum_{i=1}^{N}x_{i}^{2}$,\newline the compressible Euler
equations become,%
\begin{equation}
\left\{
\begin{array}
[c]{rl}%
\rho_{t}+V\rho_{r}+\rho V_{r}+{\normalsize \frac{N-1}{r}\rho V} &
{\normalsize =0}\\
\rho\left(  V_{t}+VV_{r}\right)  +\nabla P & {\normalsize =0}%
\end{array}
\right.  \label{gamma=1}%
\end{equation}

For the studies of the compressible Euler equations, please see \cite{CW} and
\cite{Lions}. Recently, there are some research concerning the construction of
solutions of the compressible Euler or Navier-Stokes equations by the
substitutional method \cite{Li}, \cite{LW}, \cite{Y1} and \cite{Liang}. Li and
Wang assumed the linear velocity%
\begin{equation}
u(x,t)=c(t)x
\end{equation}
and substituted it into the system to derive the dynamic system about the time
function $c(t)$ \cite{LW}. Then they used the standard argument of phase
diagram to drive the blowup or global existence of the ordinary differential
equation $c(t)$.\newline On the other hand, the separation method can be
governed to seek the radial symmetric solutions by the functional form%
\begin{equation}
\rho(r,t)=\frac{f(\frac{r}{a(t)})}{a^{N}(t)}\text{ and }V(r,t)=\frac{\dot
{a}(t)}{a(t)}r
\end{equation}
(\cite{GW}, \cite{M1}, \cite{DXY}, \cite{Li}, \cite{Y1} and \cite{Y3}).

It is natural to consider the more general linear velocity%
\begin{equation}
u(x,t)=c(t)x+b(t)
\end{equation}
to construct analytical solutions. In this article, we can combine the two
conventional approaches (substitutional method and separation method) to
derive the corresponding solutions. In fact, the main idea is to substitute
the linear velocity into the compressible Euler equations and compare the
coefficients of the different polynomial degree to deduce the corresponding
functional differential equations $(c(t),b(t),\rho^{\gamma-1}(0,t)).$ We apply
the Hubble's transformation
\begin{equation}
c(t)=\frac{\dot{a}(t)}{a(t)}%
\end{equation}
to simply the system of ordinary differential equations $(a(t),b(t),\rho
^{\gamma-1}(0,t))$. After proving the existences of the corresponding ordinary
differential equations, we can show the below theorem:

\begin{theorem}
\label{thm:1 copy(1)}There exists a family of solutions for the $1$%
-dimensional compressible  Euler equations (\ref{Euler}),
\begin{equation}
\left\{
\begin{array}
[c]{c}%
\rho^{\gamma-1}(x,t)=\max\left\{  \rho^{\gamma-1}(0,t)-\frac{\gamma-1}%
{K\gamma}\left[  \dot{b}(t)+b(t)\frac{\dot{a}(t)}{a(t)}\right]  x-\frac
{(\gamma-1)\xi}{2K\gamma a^{\gamma+1}(t)}x^{2},\text{\ }0\right\}  \text{ }\\
{\normalsize u(x,t)=}\frac{\overset{\cdot}{a}(t)}{a(t)}x+b(t)\\
\ddot{a}(t)=\frac{\xi}{a^{\gamma}(t)},\text{ }a(0)=a_{0}>0\text{, }\dot
{a}(0)=a_{1}\\
\ddot{b}(t)+\frac{(1+\gamma)\dot{a}(t)}{a(t)}\dot{b}(t)+\left[  \frac{2\xi
}{a^{\gamma+1}(t)}+\left(  \gamma-1\right)  \frac{\dot{a}^{2}(t)}{a^{2}%
(t)}\right]  b(t)=0,\text{ }b(0)=b_{0}\text{, }\dot{b}(0)=b_{1}\\
\frac{\partial}{\partial t}\rho^{\gamma-1}(0,t)+\rho^{\gamma-1}(0,t)\frac
{\dot{a}(t)}{a(t)}-\frac{\gamma-1}{K\gamma}\left[  \dot{b}(t)+b(t)\frac
{\dot{a}(t)}{a(t)}\right]  b(t)=0,\text{ }\rho(0,0)=\alpha
\end{array}
\right.  \label{solutionper}%
\end{equation}
where $a_{0}$, $a_{1}$, $b_{1}$, $b_{2}$ and $\alpha$ are arbitrary constants.
\end{theorem}

Our solutions (\ref{solutionper}) fully cover the previous known ones for the
1-dimensional case in \cite{Li} and \cite{LW} with $b_{0}=b_{1}=0$.

\section{Perturbational Method}

Our method does not rely on the phase diagram to show the blow or global
existence. The main tool is to only use the previous well known properties of
the Emden equation $a(t)$ (\ref{solutionper})$_{3}$.

\begin{proof}
First, we perturb the velocity as this form:
\begin{equation}
u(x,t)=c(t)x+b(t).
\end{equation}
The 1-dimensional momentum equation (\ref{Euler})$_{2}$ becomes for the
non-trivial solutions:%
\begin{equation}
(u_{t}+uu_{x})+K\frac{1}{\rho}\frac{\partial\rho_{{}}^{\gamma}}{\partial x}=0
\end{equation}
for $\gamma>1,$%
\begin{equation}
\dot{c}(t)x+\dot{b}(t)+[c(t)x+b(t)]c(t)+\frac{K\gamma}{\gamma-1}\frac
{\partial}{\partial x}\rho^{\gamma-1}=0
\end{equation}%
\begin{equation}
\frac{K\gamma}{\gamma-1}\frac{\partial}{\partial x}\rho^{\gamma-1}=-[\dot
{b}(t)+b(t)c(t)]-[\dot{c}(t)+c^{2}(t)]x.
\end{equation}
We take integration from $[0,x]$ to have:%
\begin{equation}
\frac{K\gamma}{\gamma-1}\int_{0}^{x}\frac{\partial}{\partial s}\rho^{\gamma
-1}ds=-[\dot{b}(t)+b(t)c(t)]\int_{0}^{x}ds-[\dot{c}(t)+c^{2}(t)]\int_{0}%
^{x}sds\label{eq123}%
\end{equation}%
\begin{equation}
\frac{K\gamma}{\gamma-1}\left[  \rho^{\gamma-1}(x,t)-\rho^{\gamma
-1}(0,t)\right]  =-[\dot{b}(t)+b(t)c(t)]x-\frac{\dot{c}(t)+c^{2}(t)}{2}x^{2}%
\end{equation}%
\begin{equation}
\rho^{\gamma-1}(x,t)=\rho^{\gamma-1}(0,t)-\frac{\gamma-1}{K\gamma}[\dot
{b}(t)+b(t)c(t)]x-\frac{\gamma-1}{2K\gamma}[\dot{c}(t)+c^{2}(t)]x^{2}%
.\label{density}%
\end{equation}
On the other hand, for the 1-dimensional mass equation (\ref{Euler})$_{1}$, we
obtain%
\begin{equation}
\rho_{t}+\rho_{x}u+\rho u_{x}=0
\end{equation}%
\begin{equation}
\rho_{t}+\left[  c(t)x+b(t)\right]  \rho_{x}+\rho c(t)=0.
\end{equation}
We multiple $\rho^{\gamma-2}$ on both sides to have%
\begin{equation}
\left(  \frac{\rho^{\gamma-1}}{\gamma-1}\right)  _{t}+\left[
c(t)x+b(t)\right]  \left(  \frac{\rho^{\gamma-1}}{\gamma-1}\right)  _{x}%
+\rho^{\gamma-1}c(t)=0.
\end{equation}
We substitute back to the above equation with equation (\ref{density}):%
\begin{equation}
\left(  \frac{\rho^{\gamma-1}}{\gamma-1}\right)  _{t}+\left[
c(t)x+b(t)\right]  \left(  \frac{\rho^{\gamma-1}}{\gamma-1}\right)  _{x}%
+\rho^{\gamma-1}c(t)
\end{equation}%
\begin{align}
&  =\frac{1}{\gamma-1}\left(  \frac{\partial}{\partial t}\rho^{\gamma
-1}(0,t)-\frac{\gamma-1}{K\gamma}\frac{\partial}{\partial t}[\dot
{b}(t)+b(t)c(t)]x-\frac{\gamma-1}{2K\gamma}\frac{\partial}{\partial t}[\dot
{c}(t)+c^{2}(t)]x^{2}\right)  \\[0.1in]
&  +\left[  c(t)x+b(t)\right]  \cdot\frac{1}{\gamma-1}\left(  -\frac{\gamma
-1}{K\gamma}\frac{\partial}{\partial x}[\dot{b}(t)+b(t)c(t)]x-\frac{\gamma
-1}{2K\gamma}\frac{\partial}{\partial x}[\dot{c}(t)+c^{2}(t)]x^{2}\right)
\\[0.1in]
&  +c(t)\left[  \rho^{\gamma-1}(0,t)-\frac{\gamma-1}{K\gamma}[\dot
{b}(t)+b(t)c(t)]x-\frac{\gamma-1}{K\gamma}\frac{[\dot{c}(t)+c^{2}(t)]}{2}%
x^{2}\right]
\end{align}%
\begin{align}
&  =\frac{1}{\gamma-1}\left(  \frac{\partial}{\partial t}\rho^{\gamma
-1}(0,t)-\frac{\gamma-1}{K\gamma}\frac{\partial}{\partial t}[\dot
{b}(t)+b(t)c(t)]x-\frac{\gamma-1}{2K\gamma}\frac{\partial}{\partial t}[\dot
{c}(t)+c^{2}(t)]x^{2}\right)  \\[0.1in]
&  +\left[  c(t)x+b(t)\right]  \left(  -\frac{1}{K\gamma}[\dot{b}%
(t)+b(t)c(t)]-\frac{1}{K\gamma}[\dot{c}(t)+c^{2}(t)]x\right)  \\[0.1in]
&  +c(t)\left[  \rho^{\gamma-1}(0,t)-\frac{\gamma-1}{K\gamma}[\dot
{b}(t)+b(t)c(t)]x-\frac{\gamma-1}{2K\gamma}[\dot{c}(t)+c^{2}(t)]x^{2}\right]
\end{align}%
\begin{align}
&  =\frac{1}{\gamma-1}\frac{\partial}{\partial t}\rho^{\gamma-1}%
(0,t)+c(t)\rho^{\gamma-1}(0,t)-\frac{b(t)}{K\gamma}[\dot{b}%
(t)+b(t)c(t)]\\[0.1in]
&  +\left\{
\begin{array}
[c]{c}%
-\frac{1}{K\gamma}\frac{\partial}{\partial t}\left[  \dot{b}%
(t)+b(t)c(t)\right]  -\frac{c(t)}{K\gamma}\left[  \dot{b}(t)+b(t)c(t)\right]
\\
-\frac{b(t)}{K\gamma}[\dot{c}(t)+c^{2}(t)]-\frac{(\gamma-1)c(t)}{K\gamma}%
[\dot{b}(t)+b(t)c(t)]
\end{array}
\right\}  x\\[0.1in]
&  +\left\{
\begin{array}
[c]{c}%
-\frac{1}{2K\gamma}\frac{\partial}{\partial t}[\dot{c}(t)+c^{2}(t)]-\frac
{c(t)}{K\gamma}[\dot{c}(t)+c^{2}(t)]\\
-\frac{(\gamma-1)c(t)}{2K\gamma}[\dot{c}(t)+c^{2}(t)]
\end{array}
\right\}  x^{2}%
\end{align}
By comparing the coefficients of the polynomial, we require the functional
differential equations involving $(c(t),b(t),\rho^{\gamma-1}(0,t))$:
\begin{equation}
\left\{
\begin{array}
[c]{c}%
\frac{d}{dt}\rho^{\gamma-1}(0,t)+(\gamma-1)c(t)\rho^{\gamma-1}(0,t)-\frac
{\gamma-1}{K\gamma}b(t)[\dot{b}(t)+b(t)c(t)]=0\\
\frac{d}{dt}[\dot{b}(t)+b(t)c(t)]+\gamma c(t)[\dot{b}(t)+b(t)c(t)]+b(t)[\dot
{c}(t)+c^{2}(t)]=0\\
\frac{d}{dt}[\dot{c}(t)+c^{2}(t)]+(\gamma+1)[\dot{c}(t)+c^{2}(t)]c(t)=0
\end{array}
\right.  \label{ODE}%
\end{equation}
to solve the 1-dimensional compressible Euler system.\newline For details
(existence, uniqueness and continuous dependence) about theories of functional
differential equations, the interested reader could see the classical
literatures \cite{H} and \cite{Wa}.\newline Here, we solve equation
(\ref{ODE})$_{3}$ with the Hubble's expression for $c(t)$:%
\begin{equation}
c(t)=\frac{\dot{a}(t)}{a(t)}%
\end{equation}%
\begin{equation}
\frac{d}{dt}\left[  \frac{d}{dt}\left(  \frac{\dot{a}(t)}{a(t)}\right)
+\frac{\dot{a}^{2}(t)}{a^{2}(t)}\right]  +(\gamma+1)\left[  \frac{d}%
{dt}\left(  \frac{\dot{a}(t)}{a(t)}\right)  +\frac{\dot{a}^{2}(t)}{a^{2}%
(t)}\right]  \frac{\dot{a}(t)}{a(t)}=0
\end{equation}%
\begin{equation}
\frac{d}{dt}\left[  \frac{\ddot{a}(t)}{a(t)}-\frac{\dot{a}^{2}(t)}{a^{2}%
(t)}+\frac{\dot{a}^{2}(t)}{a^{2}(t)}\right]  +(\gamma+1)\left[  \frac{\ddot
{a}(t)}{a(t)}-\frac{\dot{a}^{2}(t)}{a^{2}(t)}+\frac{\dot{a}^{2}(t)}{a^{2}%
(t)}\right]  \frac{\dot{a}(t)}{a(t)}=0
\end{equation}%
\begin{equation}
\left\{
\begin{array}
[c]{c}%
\frac{d}{dt}\left[  \frac{\ddot{a}(t)}{a(t)}\right]  +(\gamma+1)\frac{\ddot
{a}(t)}{a(t)}\frac{\dot{a}(t)}{a(t)}=0\\
a(0)=a_{0}>0,\text{ }\dot{a}(0)=a_{1}\text{, }\dddot{a}(0)=a_{2}%
\end{array}
\right.
\end{equation}%
\begin{equation}
\frac{\dddot{a}(t)}{a(t)}-\frac{\dot{a}(t)\ddot{a}(t)}{a^{2}(t)}%
+(\gamma+1)\frac{\dot{a}(t)\ddot{a}(t)}{a^{2}(t)}=0
\end{equation}%
\begin{equation}
\frac{\dddot{a}(t)}{a(t)}+\gamma\frac{\dot{a}(t)\ddot{a}(t)}{a^{2}(t)}=0.
\end{equation}
We multiple $a^{\gamma+1}(t)$ on both sides:%
\begin{equation}
a^{\gamma}(t)\dddot{a}(t)+\gamma a^{\gamma-1}(t)\dot{a}(t)\ddot{a}(t)=0.
\end{equation}
Here, we can observe it can reduce to the Emden equation:%
\begin{equation}
\left\{
\begin{array}
[c]{c}%
\ddot{a}(t)=\frac{\xi}{a^{\gamma}(t)}\\
a(0)=a_{0}>0\text{, }\dot{a}(0)=a_{1}%
\end{array}
\right.  \label{Emdeneq1}%
\end{equation}
where $\xi=a_{0}^{\gamma}a_{2}$ are an arbitrary constant by choosing $a_{2}%
.$\newline We remark that the Emden equation (\ref{Emdeneq1}) was well studied
in the literature of astrophysics and mathematics. The local existence of the
Emden equation (\ref{Emdeneq1}) can be promised by the fixed point theorem.

For the second equation (\ref{ODE})$_{2}$ of the dynamic system, we have%
\begin{equation}
\frac{d}{dt}\left[  \dot{b}(t)+b(t)\frac{\dot{a}(t)}{a(t)}\right]
+\gamma\left[  \dot{b}(t)+b(t)\frac{\dot{a}(t)}{a(t)}\right]  \frac{\dot
{a}(t)}{a(t)}+\frac{\ddot{a}(t)}{a(t)}b(t)=0
\end{equation}%
\begin{equation}
\ddot{b}(t)+\dot{b}(t)\frac{\dot{a}(t)}{a(t)}+b(t)\frac{d}{dt}\frac{\dot
{a}(t)}{a(t)}+\gamma\frac{\dot{a}(t)}{a(t)}\dot{b}(t)+\gamma b(t)\frac{\dot
{a}^{2}(t)}{a^{2}(t)}+\frac{\ddot{a}(t)}{a(t)}b(t)=0
\end{equation}%
\begin{equation}
\ddot{b}(t)+\frac{(1+\gamma)\dot{a}(t)}{a(t)}\dot{b}(t)+\left[  \frac{\ddot
{a}(t)}{a(t)}-\frac{\dot{a}^{2}(t)}{a^{2}(t)}+\frac{\gamma\dot{a}^{2}%
(t)}{a^{2}(t)}+\frac{\ddot{a}(t)}{a(t)}\right]  b(t)=0
\end{equation}%
\begin{equation}
\left\{
\begin{array}
[c]{c}%
\ddot{b}(t)+\frac{(1+\gamma)\dot{a}(t)}{a(t)}\dot{b}(t)+\left[  \frac{2\xi
}{a^{\gamma+1}(t)}+\left(  \gamma-1\right)  \frac{\dot{a}^{2}(t)}{a^{2}%
(t)}\right]  b(t)=0\\
b(0)=b_{0}\text{, }\dot{b}(0)=b_{1}%
\end{array}
\right.
\end{equation}
with the Emden equation (\ref{Emdeneq1}) for $a(t)$. \newline We denote
$f_{1}(t)=\frac{\left(  1+\gamma\right)  \dot{a}(t)}{a(t)}$ and $f_{2}%
(t)=\left[  \frac{2\xi}{a^{\gamma+1}(t)}+\left(  \gamma-1\right)  \frac
{\dot{a}^{2}(t)}{a^{2}(t)}\right]  $ to have%
\begin{equation}
\left\{
\begin{array}
[c]{c}%
\ddot{b}(t)+f_{1}(t)\dot{b}(t)+f_{2}(t)b(t)=0\\
b(0)=b_{0}\text{, }\dot{b}(0)=b_{1}.
\end{array}
\right.
\end{equation}
Therefore, when the functions $f_{1}(t)$ and $f_{2}(t)$ are bounded, that is
\begin{equation}
\left\vert f_{1}(t)\right\vert <F_{1}\text{ and }\left\vert f_{2}%
(t)\right\vert <F_{2}%
\end{equation}
with the constant $F_{1}$ and $F_{2}$, provided that $a(t)\neq0$ and $\dot
{a}(t)$ exist for $0\leq t<T$, the functions $b(t)$ and $\dot{b}(t)$ exist and
are bounded by the comparison theorem of ordinary differential
equations.\newline For the first equation (\ref{ODE})$_{1}$, as it is a first
order ordinary differential equations only, we can solve direct
\begin{equation}
\frac{d}{dt}\rho^{\gamma-1}(0,t)+(\gamma-1)\rho^{\gamma-1}(0,t)\frac{\dot
{a}(t)}{a(t)}-\frac{\gamma-1}{K\gamma}\left[  \dot{b}(t)+b(t)\frac{\dot{a}%
(t)}{a(t)}\right]  b(t)=0.
\end{equation}
Denote $H(t)=(\gamma-1)\frac{\dot{a}(t)}{a(t)}$ and $G(t)=\frac{\gamma
-1}{K\gamma}\left[  \dot{b}(t)+b(t)\frac{\dot{a}(t)}{a(t)}\right]  b(t)$ to
solve%
\begin{equation}
\frac{d}{dt}\rho^{\gamma-1}(0,t)+\rho^{\gamma-1}(0,t)H(t)=G(t)
\end{equation}
with the bounded $a(t)\neq0$ and $\dot{a}(t)$ for $0\leq t<T$.\newline The
formula of the first order ordinary differential equation is%
\begin{equation}
\rho^{\gamma-1}(0,t)=\frac{\int_{0}^{t}\mu(s)G(s)ds+k}{\mu(t)}%
\end{equation}
with%
\begin{equation}
\mu(t)=e^{\int_{0}^{t}H(s)ds}%
\end{equation}
and a constant $k.$\newline Therefore, we have the density function by
equation (\ref{density})%
\begin{equation}
\rho^{\gamma-1}(x,t)=\rho^{\gamma-1}(0,t)-\frac{\gamma-1}{K\gamma}\left[
\dot{b}(t)+b(t)\frac{\dot{a}(t)}{a(t)}\right]  x-\frac{\gamma-1}{2K\gamma
}\frac{\ddot{a}(t)}{a(t)}x^{2}%
\end{equation}%
\begin{equation}
\rho^{\gamma-1}(x,t)=\rho^{\gamma-1}(0,t)-\frac{\gamma-1}{K\gamma}\left[
\dot{b}(t)+b(t)\frac{\dot{a}(t)}{a(t)}\right]  x-\frac{(\gamma-1)\xi}{2K\gamma
a^{\gamma+1}(t)}x^{2}.
\end{equation}
For the non-negative density solutions $\rho(x,t)$, we must set%
\begin{equation}
\rho^{\gamma-1}(x,t)=\max\left\{  \rho^{\gamma-1}(0,t)-\frac{\gamma-1}%
{K\gamma}\left[  \dot{b}(t)+b(t)\frac{\dot{a}(t)}{a(t)}\right]  x-\frac
{(\gamma-1)\xi}{2K\gamma a^{\gamma+1}(t)}x^{2},\text{ }0\right\}  .
\label{solutions88}%
\end{equation}
The proof is completed.
\end{proof}

We notice that the above solutions are not radially symmetric for the function
$b(t)\neq0$. Therefore, the above density solutions $\rho$, cannot be obtained
by separation method of the self-similar functional, as%
\begin{equation}
\rho(x,t)\neq f(\frac{x^{2}}{a(t)})g(a(t))\text{ and }u(x,t)=\frac{\dot{a}%
(t)}{a(t)}x+b(t).
\end{equation}

On the other hand, for the 1-dimensional compressible Euler system in radially
symmetry (\ref{gamma=1}), we may replace equation (\ref{eq123}) to have the
corresponding equation by taking the integration from $[0,$ $r]:$%
\begin{equation}
\frac{K\gamma}{\gamma-1}\int_{0}^{r}\frac{\partial}{\partial s}\rho^{\gamma
-1}ds=-[\dot{b}(t)+b(t)c(t)]\int_{0}^{r}ds-[c(t)+c^{2}(t)]\int_{0}^{r}sds.
\end{equation}
It is clear to have the corresponding result in radial symmetry:

\begin{theorem}
\label{thm:1}There exists a family of solutions for the $1$-dimensional
compressible Euler equations in radial symmetry (\ref{Euler}):
\begin{equation}
\left\{
\begin{array}
[c]{c}%
\rho^{\gamma-1}(r,t)=\max\left\{  \rho^{\gamma-1}(0,t)-\frac{\gamma-1}%
{K\gamma}\left[  \dot{b}(t)+b(t)\frac{\dot{a}(t)}{a(t)}\right]  r-\frac
{(\gamma-1)\xi}{2K\gamma a^{\gamma+1}(t)}r^{2},\text{\ }0\right\}  \text{ }\\
{\normalsize u(r,t)=}\frac{\overset{\cdot}{a}(t)}{a(t)}r+b(t).
\end{array}
\right.  \label{solutionradial}%
\end{equation}

\end{theorem}

It is clear to see that the solutions (\ref{solutionper}) and
(\ref{solutionradial}) are also the solutions of the 1-dimensional
compressible Navier-Stokes equations:%
\begin{equation}
\left\{
\begin{array}
[c]{rl}%
{\normalsize \rho}_{t}{\normalsize +\nabla\cdot(\rho u)} & {\normalsize =}%
{\normalsize 0}\\
{\normalsize (\rho u)}_{t}{\normalsize +\nabla\cdot(\rho u\otimes u)+\nabla P}
& {\normalsize =\mu\Delta u}%
\end{array}
\right.
\end{equation}
with a positive constant $\mu.$

\section{Blowup or Global Solutions}

To determine if the solutions are global or local only, we could use the
following lemma about the Emden equation (\ref{solutionper})$_{3}$.

\begin{lemma}
\label{lemma33 copy(1)}For the Emden equation
\begin{equation}
\left\{
\begin{array}
[c]{c}%
\ddot{a}(t)=\frac{\xi}{a^{\kappa}(t)}\\
a(0)=a_{0}>0,\text{ }\dot{a}(0)=a_{1}%
\end{array}
\right.  \label{emden1/2}%
\end{equation}
with the constant $\kappa>1,$\newline(1) if $\xi<0$%
\begin{equation}
a_{1}<\sqrt{\frac{-2\xi}{\kappa-1}}a_{0}^{\frac{(-\kappa+1)}{2}},
\end{equation}
there exists a finite time $T$, such that
\begin{equation}
\underset{t\rightarrow T^{-}}{\lim}a(t)=0,
\end{equation}
otherwise, the solution $a(t)$ exists globally, such that%
\begin{equation}
\underset{t\rightarrow+\infty}{\lim}a(t)=+\infty.
\end{equation}
(2) if $\xi=0$, with $a_{1}<0$, the solution $a(t)$ blows up in
\begin{equation}
T=\frac{-a_{0}}{a_{1}},
\end{equation}
otherwise, the solution $a(t)$ exists globally.\newline(3) if $\xi>0$, the
solution $a(t)$ exists globally, such that%
\begin{equation}
\underset{t\rightarrow+\infty}{\lim}a(t)=+\infty.
\end{equation}

\end{lemma}

All the proof can be shown by the standard energy method of classical
mechanics. The particular proofs can be found in \cite{DXY} with $\kappa>1$
for blowup cases. Therefore, we can omit the proof here.\newline We observe
that the gradient of the velocity is
\begin{equation}
\frac{\partial}{\partial x}u(x,t)=\frac{\dot{a}(t)}{a(t)}.
\end{equation}
When the function $a(T)=0$ with a finite time $T$, $\frac{\partial}{\partial
x}u(x,T)$ blows up at every space-point $x.$ And based on the above lemma
about the Emden equation for $a(t)$, (\ref{solutionper})$_{3}$, it is clear to
have the corollary:

\begin{corollary}
(1a) For $\xi<0$ and%
\begin{equation}
a_{1}<\sqrt{\frac{-2\xi}{\kappa-1}}a_{0}^{\frac{(-\kappa+1)}{2}},
\end{equation}
the solutions (\ref{solutionper}) and (\ref{solutionradial}) blow up in a
finite time $T;$\newline(1b) For $\xi=0$, with $a_{1}<0$, the solutions
(\ref{solutionper}) blow up in
\begin{equation}
T=\frac{-a_{0}}{a_{1}}.
\end{equation}
(2) otherwise, the solutions (\ref{solutionper}) and (\ref{solutionradial})
exist globally.
\end{corollary}

We remark that we also apply this perturbational method to handle the
2-component Camassa-Holm equations
\begin{equation}
\left\{
\begin{array}
[c]{c}%
\rho_{t}+u\rho_{x}+\rho u_{x}=0,\text{ }x\in R\\
m_{t}+2u_{x}m+um_{x}+\sigma\rho\rho_{x}=0
\end{array}
\right.  \label{2com}%
\end{equation}
with
\begin{equation}
m=u-\alpha^{2}u_{xx}\label{meq}%
\end{equation}
\cite{Y4} and \cite{Yuen4}.

\end{document}